\documentclass[pra,a4paper,showpacs,superscriptaddress]{revtex4}
\usepackage{amsmath}
\usepackage{amsfonts}
\usepackage{graphicx}
\usepackage{longtable}
\newcommand{\be}{\begin{equation}}
\newcommand{\ee}{\end{equation}}
\newcommand{\bea}{\begin{eqnarray}}
\newcommand{\eea}{\end{eqnarray}}

\newcommand{\ba}[1]{\left(\begin{array}{#1}}
\newcommand{\ea}{\end{array}\right)}
\begin{document}

\title{Monogamous nature of symmetric multiqubit states with two distinct spinors} 
\author{Sudha} 
\email{arss@rediffmail.com} 
\affiliation{Department of Physics, Kuvempu University, 
Shankaraghatta-577 451, Karnataka, India}
\affiliation{Inspire Institute Inc., Alexandria, Virginia, 22303, USA.}
\author{K. S. Akhilesh }
\affiliation{Department of Studies in Physics, University of Mysore, Manasagangotri, Mysuru-570006, Karnataka, India}
\author{B. G. Divyamani} 
\affiliation{Tunga Mahavidyalaya, Thirthahalli-577432, Karnataka, India}
\author{A. R. Usha Devi}
\affiliation{Department of Physics, Bangalore University, 
Bangalore-560 056, India} 
\affiliation{Inspire Institute Inc., Alexandria, Virginia, 22303, USA.}
\author{K. S. Mallesh} 
\affiliation{Department of Studies in Physics, University of Mysore, Manasagangotri, Mysuru-570006, Karnataka, India}
\date{\today}

\begin{abstract} 
Monogamy relations place restrictions on the shareability of quantum corellations in multipartite states. Being an intrinsic quantum feature, monogamy property throws light on {\emph{residual}} entanglement, an entanglement which is not accounted for by the pairwise entanglement in the state. Expressed in terms of suitable pairwise entanglement measures such as concurrence, the monogamy inequality leads to the evaluation of {\emph{tangle}}, a measure of residual entanglement.  In this work, we explore monogamy relations in pure symmetric multiqubit states constituted by two distinct spinors, the so-called {\emph{Dicke-class}} of states. Pure symmetric $N$-qubit states constituted by permutation of two orthogonal qubits form the well-known Dicke states. Those $N$-qubit pure symmetric states constructed by permutations of two non-orthogonal qubits are a one-parameter class of generalized Dicke states. With the help of Majorana geometric representation and angular momentum algebra, we analyze the bounds on monogamy inequality, expressed in terms of  squared concurrence/squared negativity of partial transpose. We show that the states with equal distribution of the two spinors are more monogamous and hence possess larger residual entanglement when compared to other inequivalent classes with different degeneracy configurations.  
\end{abstract}
\pacs{03.65.Ud, 03.67.Bg} 
\maketitle
\section{Introduction} \section{Introduction} 
Quantum entanglement is an important resource for several quantum information processing tasks which are impossible in information processing using classically correlated states. One among the distinct properties of quantum entanglement that separates it from classical correlations, is its restricted shareability. While classical correlations are infinitely shareable, there is a limitation on the manner in which quantum entanglement is distributed among its subsystems. For instance, in a tripartite system, entanglement of one party with another limits its entanglement with the third party. This unique feature of quantum entanglement is termed `{\emph{monogamy of entanglement}}~\cite{ckw} and has evoked a lot of interest
\cite{ckw,osb,denw,ter,kw2,bs1,proof,hd,neg,hd2,newjmathphys2007,misc,22n,tripartite,bs2,misc1,misc2,
newpra2009,renyi,mixed,seevinck,cpb2010,njp,paw,bs3,gl,prabhu,sudha,bruss,eofm,xi,hf2,sqd,bai2,newpra2014,newqip,salini,pjg1,disent,
pjg2,tsallis,genwclass,pjg3,anyd,trenyi,noineq,tight,tightw,4qu,conassistw,polygon,qucorr,monqureview,z19} in the quantum information community. The importance of multipartite quantum states exhibiting monogamous nature is due to their applicability in quantum communication tasks such as secure quantum key distribution~\cite{ter,paw} and reliable quantum teleportation~\cite{z19}.  

Quantifying `{\emph{tripartite entanglement}}' or the so-called `{\emph{residual entanglement}}', which is not accounted for by the pairwise entanglement in the state, is another important issue which can be addressed using monogamous nature of composite quantum states. The restricted shareability in a multiqubit state is captured in the monogamy 
inequality 
\be
\label{moninqD}
D_{A_1:A_2A_3\cdots A_N}\geq \left( D_{A_{1}A_{2}}+ D_{A_{1}A_{3}}+D_{A_{1}A_{4}}+ \cdots+ D_{A_{1}A_{N}} \right).   
\ee
Here $D_{A_{i}A_{j}}$, $i\neq j=1,\,2,\,3\ldots N$ is a suitable measure of pairwise entanglement and 
$D_{A_1:A_2A_3\cdots A_N}$ quantifies entanglement between one party (say $A_1$) with all other parties $A_2,\,A_3,\,\ldots A_N$ taken together. 
Eq. (\ref{moninqD}) indicates that the sum of pairwise entanglements in a composite state can never exceed the entanglement between one party and the remaining parties. Quantification of three-party entanglement can be done through the non-zero quantity 
\be
\label{dcon}
D_{A_1:A_2A_3\cdots A_N}- \left( D_{A_{1}A_{2}}+ D_{A_{1}A_{3}}+D_{A_{1}A_{4}}+ \cdots+ D_{A_{1}A_{N}} \right)  
\ee  
called {\emph {tangle}} with respect to the chosen measure $D_{A_{i}A_{j}}$ of pairwise entanglement.  

Choosing squared concurrence $C$~\cite{con1,con2} as a measure of pairwise entanglement, Coffman, Kundu and Wootters~\cite{ckw} proposed $\tau_c=C^2_{A:BC}-(C^2_{AB} + C^2_{AC})$, the so-called {\emph {three-tangle}} or {\emph {concurrence-tangle}} as a measure of {\emph{residual entanglement}} in {\emph{three qubit pure states}}. Its generalization to $N$-qubit pure states has been carried out by Osborne and Verstraete~\cite{osb} and the measure of residual entanglement given by
 \be
\label{Ncon}
\tau_N=C_{A_{1}:A_{2}A_{3}A_{4} \ldots A_{N}}^{2}-\left( C_{A_{1}A_{2}}^{2} + C_{A_{1}A_{3}}^{2}+C_{A_{1}A_{4}}^{2} + \cdots+ C_{A_{1}A_{N}}^{2} \right)
\ee 
is termed the $N$-concurrence tangle~\cite{osb}. 

Monogamy inequality in terms of different measures of entanglement such as squared negativity of partial transpose~\cite{ppt1,ppt2,ppt3} and square of entanglement of formation~\cite{con1,con2} are proposed in Refs. \cite{neg} and \cite{bai2}. It is shown that~\cite{pjg2} 
W-class of states have vanishing concurrence-tangle~\cite{ckw} but non-zero negativity tangle~\cite{neg,pjg2}. Choosing a particular measure of residual entanglement, amongst the several available choices, is a non-trivial task. 
Despite this difficulty, the choice of a {\emph {single}} convenient measure for an entire class of states serves to quantify the residual entanglement, {\emph{with respect to the chosen measure}}.

Symmetric multiqubit states form an important class of states due to their theoretical significance and experimental relevance. There has been a considerable experimental progress in controlled generation of multiqubit Dicke states in physical systems like photons, cold atoms and trapped ions~\cite{2004,2009,multi2009,Ion2009,NatCom2017}. Innovative experimental schemes have also been proposed to produce a large variety of symmetric multiqubit photonic states~\cite{newexpt1,newexpt2}. Establishing an entirely non-classical feature such as restricted shareablity of quantum entanglement/monogamous nature in symmetric multiqubit states is bound to have immense impact in quantum information technology in general, and secure quatum communications in particular. 

In this paper we focus our attention on monogamy property of $N$-qubit pure symmetric states constituted by two distinct qubits. This set of states i.e., $N$-qubit pure symmetric states characterized by only two distinct qubits is defined as the {\emph{Dicke-class of states}}~\cite{spin1,spin2}. Dicke states, the common eigenstates of collective angular momentum operators $\hat{J}^2$, $\hat{J}_z$, consist of two orthogonal spinors~\cite{spin1,spin2}. Pure symmetric states of $N$ qubits, constructed by permutations of any two arbitrary non-orthogonal spinors, form a generalized class of Dicke states. Both Dicke- and generalized Dicke states are represented by two distinct points on the Bloch sphere based on Majorana's geometric description~\cite{maj,solano,bastin,usrmaj,ref2a,ref2b}. 

The paper is organized as follows: In Section 2, we employ Majorana geometric representation~\cite{maj,solano,bastin,usrmaj,ref2a,ref2b} to obtain a simplified, one-parameter form of states in the Dicke-class. Using this simplified form and with the help of well established angular momentum techniques, we obtain the structure of two-qubit and single-qubit marginals of the Dicke-class of states in Section 3. With the help of these reduced density matrices, 
we explore monogamous nature of Dicke-class of states in Section 4. 
A summary of results is given in Section 5. 

\section{Majorana geometric representation of pure symmetric $N$-qubit states with two distinct spinors} 

Ettore Majorana, in his novel 1932 paper~\cite{maj}, proposed that a pure spin $j=\frac{N}{2}$ quantum state can  be represented as a {\em symmetrized} combination of $N$ constituent spinors as follows:
\begin{equation}
\label{Maj}
\vert \Psi_{\rm sym}\rangle={\cal N}\, \sum_{P}\, \hat{P}\, \{\vert \epsilon_1, \epsilon_2, 
\ldots  \epsilon_N \rangle\}, 
\end{equation}  
where 
\begin{equation}
\label{spinor}
\vert\epsilon_l\rangle= \left(
\cos(\alpha_l/2)\, \vert 0\rangle +
\sin(\alpha_l/2) \, \vert 1\rangle\right) e^{i\beta_l/2},\ \ l=1,\,2,\ldots,\,N 
\end{equation}
denote arbitrary states of spinors (qubits) $\vert\epsilon_l\rangle$. 
In Eq. (\ref{Maj}), the symbol $\hat{P}$ corresponds to the set of all $N!$ permutations of the qubits and ${\cal N}$ corresponds to an overall normalization factor. 
The name Majorana {\emph {geometric}} representation is owing to the fact that it leads to an intrinsic picture of the  state in terms of 
$N$ points on the unit sphere.  The spinors $\vert \epsilon_l\rangle$, $l=1,2,\ldots, N$ of Eq. (\ref{spinor}) correspond geometrically to $N$ points on the Bloch sphere $S^2$, with the pair of angles $(\alpha_l,\beta_l)$ determining the orientation of each point on the sphere.  
Pure symmetric $N$ qubit states consisting of two distinct qubits (orthogonal as well as non-orthogonal)  
are given by~\cite{solano,bastin,usrmaj}, 
\be
\label{dnk}
\vert D_{N-k, k}\rangle = {\cal N}\, \sum_{P}\, \hat{P}\,\{ \vert \underbrace{\epsilon_1, \epsilon_1,
\ldots , \epsilon_1}_{N-k};\ \underbrace{\epsilon_2, \epsilon_2,\ldots , \epsilon_2}_{k}\rangle\}, \ \ k=1,\,2,\,3,\ldots \left[\frac{N}{2}\right] 
\ee 
It may be noted that in Eq. (\ref{dnk}), one of the spinors say $\vert \epsilon_1 \rangle$ occurs $N-k$ times whereas the  other spinor $\vert \epsilon_2 \rangle$ occurs $k$ times in each term of the symmetrized combination ($\left[\frac{N}{2}\right]=\frac{N}{2}$ when $N$ is even and $\left[\frac{N}{2}\right]=\frac{N-1}{2}$ when $N$ is odd).  
It has been shown in Refs.~\cite{sudha,usrmaj} that the states $\vert D_{N-k, k}\rangle$ are equivalent, under identical\footnote{For any symmetric state to be transformed into another symmetric state through local unitary transformations, the unitary transformations are to be identical in order to retain the symmetry of the state (See Ref.~\cite{usrmaj}).} local unitary transformations, to a canonical form, characterized by only one real parameter. More specifically, pure symmetric multiqubit states $\vert D_{N-k, k}\rangle$ in Eq.(\ref{dnk}) can be brought~\cite{sudha} to the form
\begin{eqnarray}
\label{nono}
\vert D_{N-k, k}\rangle &\equiv & \sum_{r=0}^k\, \beta^{(k)}_{r}\,\,  \left\vert\frac{N}{2},\frac{N}{2}-r \right\rangle,  \\ 
\beta^{(k)}_{r}&=&{\cal N}\,\, 
\sqrt{\frac{N!(N-r)!}{r!}}\,\frac{a^{k-r}\, b^r}{(N-k)! (k-r)!}, \ \ \ 0<a<1,  \ b=\sqrt{1-a^2}. \nonumber 
 \end{eqnarray}
Here $\left\vert\frac{N}{2},\frac{N}{2}-r \right\rangle$, $r=0,\,1,\,2\ldots N$ denote the Dicke states, which are common eigenstates of the collective angular momentum operators $\hat{J}^2$, $\hat{J}_z$ and are the basis states of the symmetric subspace of collective angular momentum space,  with dimension $N+1$. 
The generalized Dicke states $\vert D_{N-k, k}\rangle$ are characterized by only one real parameter `$a$' and they belong to the one parameter family, the Dicke-class of states. 
The Dicke-class consists of both Dicke-states (when $a=0$) as well as generalized Dicke states (when $0<a<1$). While the Dicke states having parameter $a=0$ are characterized by two {\emph {orthogonal}} spinors $\vert 0\rangle$, $\vert 1\rangle$, the states $\vert D_{N-k,\, k}\rangle$ with $0<a<1$ are characterized by two {\emph {non-orthogonal}} spinors $\vert \epsilon_1\rangle$, $\vert \epsilon_2\rangle$. The parameter `$a$' can thus be termed as {\emph{non-orthogonality parameter}}.  

It is important to notice that in the Dicke-class of states $\vert D_{N-k,\,k}\rangle$, different values of 
$k$ ($k=1,\,2,\,3,\ldots \left[\frac{N}{2}\right]$), correspond to SLOCC inequivalent classes~\cite{solano,bastin,usrmaj,dur}.   In fact, different values of $k$ lead to different {\emph {degeneracy configurations}}~\cite{solano,bastin,usrmaj} of the two spinors. For instance, when $N=4$, there are two possible degeneracy configurations\footnote{In $\vert D_{2, 2}\rangle$, the spinors $\vert \epsilon_1 \rangle$, $\vert\epsilon_2 \rangle$ appear two times in each term of the symmetrized combination shown in Eq. (\ref{dnk}). That is, 
\[\vert D_{2, 2}\rangle={\cal{N}}\left[\vert \epsilon_1\epsilon_1\epsilon_2\epsilon_2\rangle+\vert\epsilon_1\epsilon_2\epsilon_1\epsilon_2\rangle+\vert\epsilon_1\epsilon_2\epsilon_2\epsilon_1\rangle+\vert \epsilon_2\epsilon_1\epsilon_2\epsilon_1\rangle+\vert\epsilon_2\epsilon_2\epsilon_1\epsilon_1\rangle\right].\] 
Similarly  
\[
\vert D_{3, 1}\rangle={\cal{N}}\left[\vert \epsilon_1\epsilon_1\epsilon_1\epsilon_2\rangle+\vert\epsilon_1\epsilon_1\epsilon_2\epsilon_1\rangle+\vert\epsilon_1\epsilon_2\epsilon_1\epsilon_1\rangle+\vert\epsilon_2\epsilon_1\epsilon_1\epsilon_1\rangle\right].\]
Here, spinor $\vert \epsilon_1 \rangle$ appears three times whereas $\vert \epsilon_2 \rangle$ appears only once in each term of the symmetrized combination.} corresponding to 
$k=3$ and $k=2$. 
The states $\vert D_{2, 2}\rangle$ belong to the class $\{{\cal{D}}_{2,\,2}\}$ and the states $\vert D_{3, 1}\rangle$ belong to 
$\{{\cal{D}}_{3,\,1}\}$. Both $\{{\cal{D}}_{2,\,2}\}$ and $\{{\cal{D}}_{3,\,1}\}$ are the only possible classes when $N=4$. 
In general, for any $N$, the Dicke-class of states 
is a collection of all inequivalent classes $\{{\cal{D}}_{N,\,k}\}$, $k=1,\,2,\,3,\ldots \left[\frac{N}{2}\right]$.  
In the next section, we evaluate the two-qubit and single qubit marginal density matrices of the state $\vert D_{N-k, k}\rangle$ for different values of  $k=1,\,2,\,3,\ldots \left[\frac{N}{2}\right]$. We then proceed to analyze the monogamy property of the states belonging to different inequivalent classes $\{{\cal{D}}_{N,\,k}\}$  with respect to two well-known measures of two-qubit entanglement.  

\section{Single and two-qubit reduced density matrices of generalised Dicke states} 
The monogamy relation in Eq. (\ref{moninqD}) requires evaluation of measures of entanglement between any two qubits and also between a given qubit with all other qubits in the state. For quantification of pairwise entanglement through any suitable measure of entanglement, one needs to evaluate the two-qubit reduced density matrix of the multiqubit state under consideration. Towards this end, 
we proceed to evaluate single and two qubit reduced density matrices of pure symmetric states belonging to Dicke-class..
For the pure symmetric multiqubit state  $\vert D_{N-k,\,k}\rangle$, owing to exchange symmetry, all reduced density matrices are identical. Thus, the two-qubit marginal density matrix $\rho^{(k)}_2$ corresponding to {\emph{any}} random pair of qubits in the $N$-qubit symmetric state $\vert D_{N-k,\,k}\rangle$  is obtained by tracing over the remaining $N-2$ qubits in it. 

On using the form of the state 
$\vert D_{N-k,\,k}\rangle$ given in  Eq. ({\ref{nono}}), we have 
\begin{eqnarray}
\label{formal_rhok1}
\rho^{(k)}_2&=&\mbox{Tr}_{N-2}\,\left(\vert D_{N-k,\,k}\rangle \langle D_{N-k,\,k}\vert\right),\ \ \  k=1,\,2,\,3,\ldots \left[\frac{N}{2}\right];\nonumber \\
&=&\mbox{Tr}_{N-2}\left(\sum_{r,r'=0}^k\, \beta^{(k)}_{r}\, \beta^{(k)}_{r'}\,  \left\vert\frac{N}{2},\frac{N}{2}-r \right\rangle
 \left\langle\frac{N}{2},\frac{N}{2}-r' \right\vert \right).
\end{eqnarray}
To facilitate the tracing operation over $N-2$ qubits, we partition the state $\vert D_{N-k,\,k}\rangle$ into $N-2$ qubits and two qubits. As a pure symmetric state with $N-2$ qubits is equivalent to an angular momentum state
$\vert j_1,\,m_1\rangle$ with $j_1=\frac{N-2}{2}$
and a two qubit pure symmetric state is equivalent to the state $\vert j_2,\,m_2\rangle$ with  $j_2=1$, we use the addition of angular momenta relation~\cite{Var}  
\be
\label{angj1j2}
\vert j,\,m\rangle =\sum_{m_2=-j_2}^{j_2}\, C(j_1,\, j_2,\, j;m-m_2,\, m_2,\, m) \, \left(\vert j_1,m-m_2 \rangle\otimes\vert j_2,m_2 \rangle\right)
\ee 
for the required partition of the $N$-qubit state $\vert D_{N-k,\,k}\rangle$. Here $C(j_1,\, j_2,\, j;m-m_2,\, m_2, m)$ are the Clebsch-Gordan coefficients~\cite{Var} in the addition of angular momenta, with quantum numbers $j_1$ and $j_2$.
With $j_1=\frac{N}{2}-1$, $j_2=1$, we obtain
\be
\label{angj1j2N2}
\left\vert\frac{N}{2},\frac{N}{2}-r \right\rangle=\sum_{m_2=-1,0,1}\,\left[ c_{m_2}^{(r)}\,\, 
\left\vert\frac{N}{2}-1,\frac{N}{2}-r-m_2 \right\rangle\otimes\vert 1,\,m_2\rangle \right], \ \ \ r=0,\,1,\ldots,N,
\ee 
where we have denoted
\begin{eqnarray}
\label{cgp}
c_{m_2}^{(r)}&=&C\left(\frac{N}{2}-1,\, 1,\, \frac{N}{2}; m-m_2,\, m_2, m\right), \\ 
& & m=-\frac{N}{2},\,-\frac{N}{2}+1,\,\ldots,\frac{N}{2},
\ \    m_2=-1,\,0,\,1,\ \   r=0,\,1,\ldots,N. \nonumber
\end{eqnarray}   
 We thus obtain (See Eqs. (\ref{formal_rhok1}) and $(\ref{angj1j2N2})$)
\begin{eqnarray*}
\rho^{(k)}_2&=&\mbox{Tr}_{N-2}\,\left\{\sum_{r,r'=0}^k\, \beta^{(k)}_r\, \beta^{(k)}_{r'} \sum_{m_2,m_2'}\,\left[ c_{m_2}^{(r)}\, 
c_{m'_2}^{(r')}\,
\left\vert\frac{N}{2}-1,\frac{N}{2}-r-m_2 \right\rangle \left\langle \frac{N}{2}-1,\frac{N}{2}-r'-m_2'   \right\vert \right. \right. \\
 & & \left. \left. \otimes \vert 1, m_2\rangle \langle 1, m_2'\vert \right]  \right\}. 
\end{eqnarray*}
More explicitly, tracing over $N-2$ qubits leads us to 
\begin{eqnarray}
\label{N-2k}
\rho^{(k)}_2&=&\left\{\sum_{r,r'=0}^k\, \beta^{(k)}_r\, \beta^{(k)}_{r'} \sum_{m_2,m_2'}\,c_{m_2}^{(r)}\, 
c_{m'_2}^{(r')}\,
\left[\left\langle\frac{N}{2}-1,\frac{N}{2}-r-m_2 \right\vert\left.\frac{N}{2}-1,\frac{N}{2}-r'-m_2'\right\rangle \right] \vert 1, m_2\rangle \langle 1, m_2'\vert   \right\} \nonumber \\
&=&\left\{\sum_{r,r'=0}^k\,\sum_{m_2,m_2'} \beta^{(k)}_r\, \beta^{(k)}_{r'} \,c_{m_2}^{(r)}\, 
c_{m'_2}^{(r')}\,\sum_{m_1=(-N/2)+1}^{(N/2)-1}\,\, \left\langle \frac{N}{2}-1, m_1    \right\vert \left.\frac{N}{2}-1,\frac{N}{2}-r'-m'_2 \right\rangle \right.\, \nonumber \\ 
& & \left. \left. \left\langle \frac{N}{2}-1,\frac{N}{2}-r-m_2 \right\vert \frac{N}{2}-1, m_1 \right\rangle 
\vert 1, m_2\rangle \langle 1, m_2'\vert   \right\}.
\end{eqnarray}
In the second line of Eq. (\ref{N-2k}), we have made use of the completeness relation 
\[
I=\sum_{m_1=(-N/2)+1}^{(N/2)-1}\, \left\vert \frac{N}{2}-1,\,m_1 \right\rangle\,\left\langle \frac{N}{2}-1,\,m_1 \right\vert,
\]  
with $I$ being the identity matrix in the space of $N-2$ qubits. We finally obtain the two qubit reduced density matrix of any random pair of qubits of the $N$-qubit state $\vert D_{N-k,\,k}\rangle$. The two-qubit reduced density matrix $\rho^{(k)}_2$ now turns out to be 
\be
\label{finalrhok}
\rho^{(k)}_2=\sum_{m_2,m_2'=1,0,-1}\, \rho^{(k)}_{m_2,m_2'}\,  \vert 1, m_2\rangle \langle 1, m_2'\vert,  
\ee
where 
\begin{eqnarray}
\label{rhok}
\rho^{(k)}_{m_2,m_2'}&=&\sum_{r,r'=0}^k\, \beta^{(k)}_r\, \beta^{(k)}_{r'}\, c_{m_2}^{(r)}\,c_{m'_2}^{(r')}\,\left\{ \sum_{m_1=(-N/2)+1}^{(N/2)-1}\, \left\langle \frac{N}{2}-1, m_1    \right\vert \left.\frac{N}{2}-1,\frac{N}{2}-r'-m'_2 \right\rangle \right. \nonumber \\
 & & \left.
\left\langle \frac{N}{2}-1,\frac{N}{2}-r-m_2   \right\vert \left. \frac{N}{2}-1, m_1\right\rangle \right\}
\end{eqnarray} 
denote the matrix elements of $\rho^{(k)}_2$ in the basis $\{\vert 1,\,m_2\rangle,\, m_2=-1,\,0,\,1\}$. 
The associated Clebsch-Gordan coefficients $c^{(r)}_{m_2}$ are explicitly given by~\cite{Var}
\begin{eqnarray}
\label{cg_explicit}
c^{(r)}_{1}&=&\sqrt{\frac{(N-r)(N-r-1)}{N(N-1)}},\ \ \ c^{(r)}_{-1}=\sqrt{\frac{r\, (r-1)}{N(N-1)}},\nonumber \\
c^{(r)}_{0}&=&\sqrt{\frac{2r\, (N-r)}{N(N-1)}}.	  
\end{eqnarray} 
Expressing the spin-$1$ states in terms of the constituent two qubit states i.e., 
\[ 
\vert 1,1\rangle=\vert 0_A,0_B\rangle,\ \ \ \ \vert 1,0\rangle=(\vert 0_A,1_B\rangle+\vert 1_A,0_B\rangle)/\sqrt{2},\ \ \ \ 
\vert 1,-1\rangle=\vert 1_A,1_B\rangle, 
\] 
the following simplified form~\cite{s2,Vidal06} is realised for the symmetric two-qubit reduced density matrix:     
\begin{eqnarray}
\label{rhok_matrix}
\rho^{(k)}_2&=&\ba{cccc} A^{(k)} \ \ & B^{(k)} \ \ & B^{(k)}\ \  & C^{(k)} \ \  \\ B^{(k)} \ \  & D^{(k)}\ \  & D^{(k)}\ \  & E^{(k)} \ \   \\ B^{(k)} \ \  & D^{(k)} \ \  & D^{(k)} \ \ & E^{(k)} \ \  \\ C^{(k)} \ \  & E^{(k)} \ \  & E^{(k)} \ \  & F^{(k)} \ \  \ea. 
\end{eqnarray}
The elements $A^{(k)},\, B^{(k)},\, C^{(k)},\, D^{(k)},\, E^{(k)}$ and $F^{(k)}$ are real and are explicitly given by~\cite{spin1}
\begin{eqnarray}
\label{elements}
A^{(k)}=\sum_{r=0}^k\, \left({\beta^{k}}_r\right)^2 \left({c^{(r)}_{1}}\right)^2, \  & & \  B^{(k)}=\frac{1}{\sqrt{2}}\sum_{r=0}^{k-1}\, 
{\beta^{(k)}_r} \beta^{(k)}_{r+1}\, c^{(r)}_{1} 
c^{(r+1)}_{0} \nonumber \\ 
& & \nonumber \\ 
C^{(k)}=\sum_{r=0}^{k-2}\, \beta^{(k)}_r \beta^{(k)}_{r+2}\,\, c^{(r)}_{1} c^{(r+2)}_{-1}, \  & & \  
D^{(k)}=\frac{1}{2}\sum_{r=1}^{k}\, \left({\beta_r^{(k)}}\right)^2  \left({c^{(r)}_{0}}\right)^2 \\
& & \nonumber \\ 
E^{(k)}=\frac{1}{\sqrt{2}}\sum_{r=0}^{k-1}\, \beta^{(k)}_r \beta^{(k)}_{r+1}\,\, c^{(r)}_{0}c^{(r+1)}_{-1}, \  & &  \  
 \ \ \ F^{(k)}=\sum_{r=0}^k\, \left({\beta_r^{(k)}}\right)^2  \left({c^{(r)}_{-1}}\right)^2 \nonumber
\end{eqnarray} 

In order to evaluate the entanglement between a single qubit say $A_1$ and the remaining qubits $A_2 A_3 \cdots A_N$, 
we recall that when the $N$-qubit state is pure, the partition containing $N-1$ qubits can be treated as a single qubit~\cite{ckw}. This observation allows any measure of pairwise entanglement suitable for quantifying entanglement between one qubit and the remaining qubits in a {\emph{pure}} multiqubit state. 
In particular, when the measure of entanglement is either concurrence~\cite{con1} or negativity of partial transpose~\cite{ppt1,ppt2,ppt3}, it is seen that~\cite{ckw,neg}
\be
\label{squbit}
C_{A_1:A_2A_3\cdots A_n}=N_{A_1:A_2A_3\cdots A_n}=2 \,\sqrt{\mbox{det}\,\rho_{A_1}},
\ee
where $\rho_{A_1}$ is the density matrix of the qubit $A_1$. 

The state $\vert D_{N-k,\,k}\rangle$  being symmetric, the marginal $\rho^{(k)}_1$ is independent of the choice of qubit. We obtain the single qubit marginal $\rho^{(k)}_1$ either by tracing $N-1$ qubits in the state $\vert D_{N-k,\,k}\rangle$ or by tracing a single qubit from the two-qubit reduced density matrix $\rho^{(k)}_2$ (See Eq.(\ref{rhok_matrix})). The reduced density matrix $\rho^{(k)}_1$ characterizing any single qubit of the pure symmetric $N$ qubit state $\vert D_{N-k},\, k\rangle$ is thus obtained as
\be
\label{rhok1_matrix}
\rho^{(k)}_1=\ba{cc}  A^{(k)}+D^{(k)}   & B^{(k)}+E^{(k)}  \\  B^{(k)}+E^{(k)}  &  D^{(k)}+F^{(k)} 
\ea.
\ee 

The two-qubit and single qubit marginals $\rho^{(k)}_2$, $\rho^{(k)}_1$ obtained explicitly through relations 
(\ref{rhok_matrix}),  (\ref{rhok1_matrix}), help in setting up the monogamy relations with respect to  any suitable measure of pairwise entanglement. In the following section, we analyse the bounds on monogamy relations for the state $\vert D_{N-k,\,k}\rangle$, choosing concurrence and negativity of partial transpose as measures of pairwise entanglement.

\section{Monogamy relation for Dicke class of states}

Having obtained the form of single and two-qubit reduced density matrices  $\rho^{(k)}_1$ and $\rho^{(k)}_2$ (See Eqs. (\ref{rhok_matrix}), 
(\ref{rhok1_matrix})) of the state $\vert D_{N-k, k}\rangle$, we will use them here to set up the monogamy relation for the Dicke-class of states. For this purpose, we choose squared concurrence and squared negativity of partial transpose as two different, yet  suitable measures of entanglement. As the Dicke class of states contain several inequivalent classes $\{{\cal{D}}_{N,\,k}\}$, with different degeneracy configuration~\cite{solano,bastin,usrmaj} of the two spinors (corresponding to $k=1,\,2,\cdots \ [N/2]$),  
we analyze the bound on monogamy relation in each class. 
\subsection{Monogamy relation for generalized Dicke states in terms of squared concurrence:} 

We recall here that the monogamy relation in terms of squared concurrence was set up for three-qubit pure states by Coffman, Kundu and 
Wootters~\cite{ckw}. It was generalized to $N$-qubit pure states in Ref.~\cite{osb}. Concurrence is a convenient measure of pairwise entanglement in two-qubit states (pure as well as mixed). For the two-qubit state $\rho^{(k)}_2$, concurrence $C$ is given by 
\[
C_{k_{2}}=\mbox{max}\left(0,\,\sqrt{\lambda_1}-\sqrt{\lambda_2}-\sqrt{\lambda_3}-\sqrt{\lambda_4} \right).
\] 
Here $\lambda_i$, $i=1,\,2,\,3,\,4$ are the  eigenvalues of the  matrix $\rho^{(k)}_2{\rho'^{(k)}_2}$ with 
${\rho'^{(k)}_2}=({\sigma}_{y}\otimes{\sigma}_{y}) \rho^{*}_2{^{(k)}}$ $({\sigma}_{y}\otimes{\sigma}_{y})$ being the spin-flipped density matrix~\cite{con1,con2}.  

With the explicit form of $\rho^{(k)}_2$ obtained in Section~3 (See Eq. (\ref{rhok_matrix})), the concurrence $C_{k_{2}}$ can readily be evaluated. 

While the concurrence $C_{k_{2}}$ quantifies the pairwise entanglement between any two qubits of the state $\vert D_{N-k,\, k}\rangle$, the entanglement  between {\emph {any}} qubit with the remaining $N-1$ qubits is given by (See Eq. (\ref{squbit}))
\be
\label{squbit2}
C_{A_1:A_2A_3\cdots A_N}\equiv C_{k_{1}}=2 \,\sqrt{\mbox{det}\,\rho^{(k)}_1}
\ee 
Notice here that, due to the symmetry of the state  $\vert D_{N-k,\, k}\rangle$ under permutation of qubits, the choice of qubit $A_1$ in 
$C_{A_1:A_2A_3\cdots A_N}\equiv C_{k_{1}}$ is arbitrary and hence Eq. (\ref{squbit2}) gives the entanglement between {\emph {any}} qubit with remaining $N-1$ qubits. With the single qubit marginal $\rho^{(k)}_1$ of the state $\vert D_{N-k,\, k}\rangle$  
being explicitly given in Section~3 (See Eq.(\ref{rhok1_matrix}) along with Eqs.,(\ref{cg_explicit}), (\ref{elements})), one can readily evaluate the concurrence $C_{k_{1}}$ (See Eq. (\ref{squbit2})) which quantifies the entanglement between any qubit with the remaining qubits. 
With reference to an arbitrary qubit of the $N$-qubit pure symmetric state $\vert D_{N-k,\, k}\rangle$, there are $N-1$ pairs of qubits (See Eqs. (\ref{dcon}), (\ref{Ncon})) and entanglement between them is quantified through squared concurrence $C^2_{k_{2}}$. The monogamy inequality for the Dicke-class of states (See Eq. (\ref{Ncon})) is therefore, given by
\be 
C^2_{k_{1}}>(N-1)C^2_{k_{2}}. 
\ee 
The bound on monogamy inequality can be ascertained by evaluating the {\emph{concurrence tangle}}
$\tau_c$ which quantifies residual entanglement (as, entanglement of $N$ qubit states which is not captured by  pairwise entanglement measures alone) 
in terms of squared concurrence~\cite{ckw}. For the states $\vert D_{N-k,\, k}\rangle$ belonging to the Dicke-class, the concurrence tangle is given by 
\be
\label{ctangle}
\tau^{(k)}_N=C^2_{k_{1}}-(N-1)C^2_{k_{2}}. 
\ee
It may be noted that larger value of  residual entanglement $\tau^{(k)}_N$ indicates that the $N$-qubit generalized Dicke states 
$\vert D_{N-k,\,k}\rangle$ are more monogamous. In other words, Eq. (\ref{ctangle}) is useful to verify constrained shareability of entanglement amongst the $N$ qubits because of monogamy property  of quantum entanglement.
We have explicitly evaluated the concurrence-tangle $\tau^{(k)}_N$ for the states belonging to inequivalent classes  
$\{{\cal {D}}_{N-k,\,k}\}$,  $k=1,\,2,\,3,\,4,\,5$ as a function of $N$ and the parameter `$a$'.  
In Fig. 1 to Fig. 4, the variation of concurrence tangle  $\tau_N^{(k)}$ (measure of residual entanglement in terms of squared concurrence) with the parameter `$a$'  is shown.

\begin{figure}[ht]
\includegraphics* [width=4in,keepaspectratio]{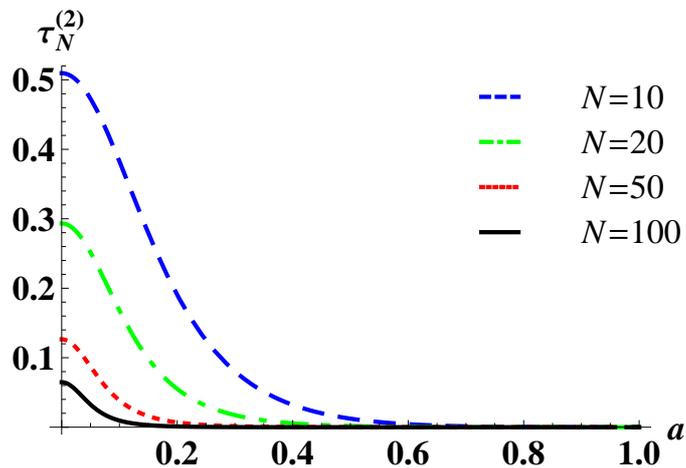} 
\caption{The plot of concurrence tangle $\tau^{(2)}_N$ as a function of the real parameter `$a$' ($0\leq a \leq 1$), for  the states 
$\vert D_{N-2,\, 2}\rangle$} 
\end{figure} 
\begin{figure}[ht]
\includegraphics*[width=4in,keepaspectratio]{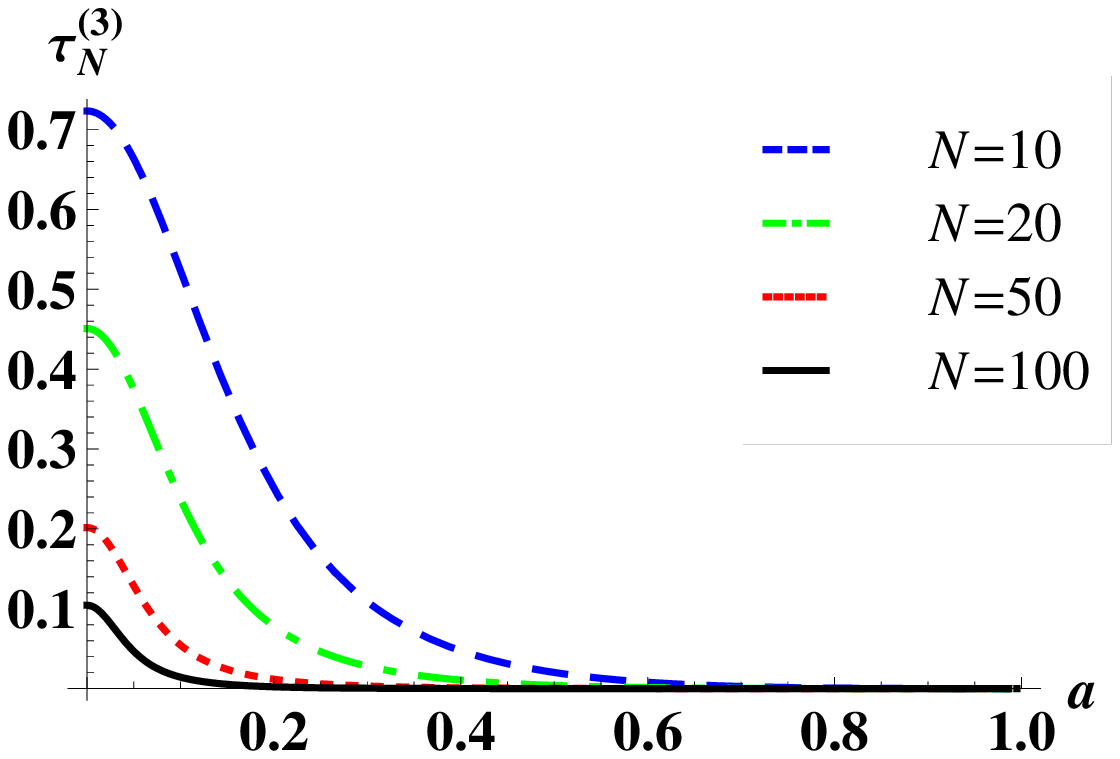} 
\caption{The plot of concurrence tangle $\tau^{(3)}_N$ as a function of the real parameter `$a$' ($0\leq a \leq 1$), for  the states 
$\vert D_{N-3,\, 3}\rangle$}
\end{figure} 
\begin{figure}[ht]
\includegraphics*[width=4in,keepaspectratio]{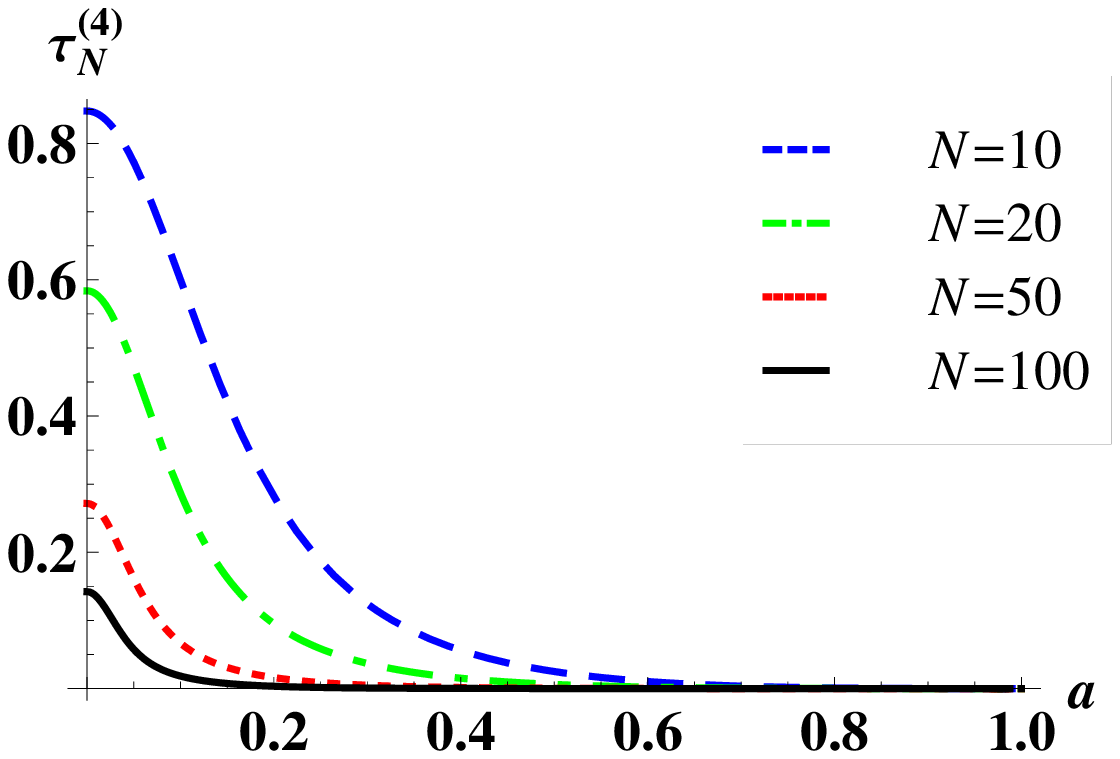} 
\caption{The plot of concurrence tangle $\tau^{(4)}_N$  as a function of the real parameter `$a$' for the states $\vert D_{N-4,\, 4}\rangle$}  
\end{figure} 

\begin{figure}[ht]
\includegraphics*[width=4in,keepaspectratio]{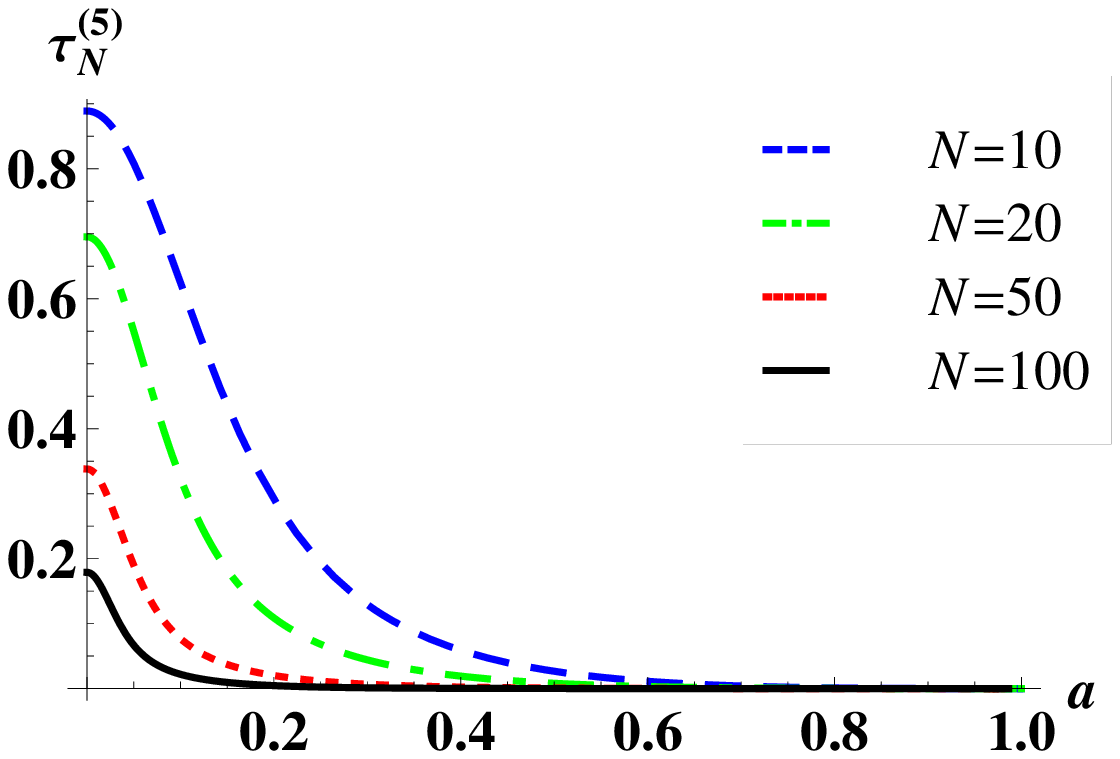} 
\caption{The plot of concurrence tangle $\tau^{(5)}_N$ as a function of the real parameter `$a$' ($0\leq a \leq 1$), for  the states 
$\vert D_{N-5,\, 5}\rangle$,}  
\end{figure} 

The states $\vert D_{N-1,\, 1}\rangle$ belong to the W-class of states~\cite{pjg2} $\{{\cal {D}}_{N-1,\,1}\}$ in which, one of the spinors appears only once in each term of the symmetrized combination (See Eq. (\ref{nono})). The well-known W states 
$\vert \frac{N}{2},\, \frac{N}{2}-1\rangle$ belong to this family and correspond to the parameter $a=0$ of the state $\vert D_{N-1,\,1}\rangle$(See Eq. (\ref{nono})). 
It has been shown in Ref.~\cite{pjg2} that the W-class of states saturate the {\emph{monogamy inequality in terms of squared concurrence}}, i.e.,
\[
C^2_{k_{1}}=(N-1)C^2_{k_{2}} \Longrightarrow \tau^{(k)}_N=C^2_{k_{1}}-(N-1)C^2_{k_{2}}=0
\]
for W-class of states.
This has the physical implication that the $N$-qubit pure state $\vert D_{N-1,\, 1}\rangle$ possesses only pairwise entanglement, when squared concurrence is chosen as measure of two-qubit entanglement. It also means that entanglement between one-qubit with the remaining qubits is equally shared amongst the pairwise entanglement in the $N-1$ pairs of qubits, when the entanglement is measured in terms of squared concurrence. We will illustrate in the next section that, when expressed in terms of squared negativity of partial transpose, the W-class of states are shown to have non-zero residual entanglement. 

From Figs. 1 to 4, one can draw the following conclusions about the bound on monogamy relation, for Dicke-class of states. 
\begin{enumerate}
\item For the {\emph{Dicke states}} $\vert \frac{N}{2},\, \frac{N}{2}-r\rangle$, $r=1,\,2,\,\cdots \,N$, characterized by two orthogonal spinors the bound on monogamy is larger compared to their companion states with non-orthogonal spinors. Figs. 1 to 4 readily illustrate that the concurrence tangle $\tau_N^{(k)}$ is maximum when the parameter
$a=0$ and monotonically decreases when $a$ ($0<a<1$) increases. For  separable states corresponding to $a=1$, the concurrence tangle vanishes, as expected. 

\item The concurrence tangle for the states $\vert D_{N-k,\,k}\rangle$ increase with the increase in the value of $k$ $(k=2,\,3,\,4,\,5)$. 
In particular, the state $\vert D_{N-k,\, k}\rangle$ with $k=\left[\frac{N}{2}\right]$,
is more monogamous. This implies that the Dicke-class of states $\vert D_{N-k,\,k}\rangle$ with equal distribution of two spinors, has larger bound on the monogamy relation and possesses larger residual entanglement. 

\item 
The residual entanglement (and hence the bound on monogamy relation with respect to squared concurrence) reduces with the increase in number of qubits $N$,  as expected.  

\end{enumerate} 

\subsection{Monogamy relation for generalized Dicke states in terms of squared negativity of partial transpose:} 

Monogamous nature of pure three qubit states with respect to negativity of partial transpose was examined in Ref.~\cite{neg}. This work gave an indication that any {\emph{single}} measure of entanglement serves to provide a bound on monogamy inequality, which is specific to that measure and the state under consideration.
For instance, it is shown in  Ref.~\cite{neg} that three qubit W-state and its non-symmetric generalization have non-zero bound on monogamy inequality, set up in terms of negativity of partial transpose. While concurrence-tangle of W- and non-symmetric generalization of W-states vanish, their negativity tangle is non-zero. It is also shown in Ref. \cite{pjg2} that for the  one-parameter family 
$\{{\cal D}_{N-1,\,1}\}$ of W-class of states $\vert D_{N-1,\,1}\rangle$ (generalized W states), consisting of two orthogonal/non-orthogonal qubits, concurrence tangle is zero but  negativity tangle has a maximum value for W-states $\vert \frac{N}{2},\,{\frac{N}{2}}-1\rangle$ and decreases monotonically with parameter 
$a$ ($0<a\leq 1$) of $\vert D_{N-1,\,1}\rangle$. 
We thus focus on investigating monogamous nature of Dicke-class of states with respect to squared negativity of partial transpose. 
%

With the knowledge of two-qubit reduced density matrix (See Eq. (\ref{rhok_matrix})) of the state $\vert D_{N-k,\,k}\rangle$, we can readily evaluate its negativity of partial transpose~\cite{ppt1,ppt2,ppt3}. The partially transposed density matrix ${\left(\rho^{(k)}_2\right)}^T$ of the two-qubit state $\rho^{(k)}_2$ (See Eq. (\ref{rhok_matrix})) is explicitly given by
\begin{eqnarray}
\label{rho2pt}
{\left(\rho^{(k)}_2\right)}^T&=&\ba{cccc} A^{(k)} \ \ & B^{(k)} \ \ & B^{(k)}\ \  & D^{(k)} \ \  \\ B^{(k)} \ \  & D^{(k)}\ \  & C^{(k)}\ \  & E^{(k)} \ \   \\ B^{(k)} \ \  & C^{(k)} \ \  & D^{(k)} \ \ & E^{(k)} \ \  \\ D^{(k)} \ \  & E^{(k)} \ \  & E^{(k)} \ \  & F^{(k)} \ \  \ea,
\end{eqnarray} 
as~\cite{ppt1} ${\left(\rho^{(k)}_2\right)}^T_{ij;kl}={(\rho^{(k)}_2)}_{il;kj}={(\rho^{(k)}_2)}_{kj;il}$
are the elements of ${\left(\rho^{(k)}_2\right)}^T$. 

The negativity of partial transpose~\cite{ppt1,ppt2,ppt3}  
is defined as
\[
N_{k_2}=\left(\left\vert\left\vert {\left(\rho^{(k)}_2\right)}^T \right\vert \right\vert-1\right)/2,
\] 
where 
$\left\vert\left\vert {\left(\rho^{(k)}_2\right)}^T\right\vert\right\vert$ is the trace-norm\footnote{The trace-norm~\cite{ppt3} $\left\vert\left\vert {\left(\rho^{(k)}_2\right)}^T\right\vert \right\vert$ is the sum of the square-roots of eigenvalues of the positive-definite matrix $\left({{(\rho^{(k)}_2)}^T}\right)^\dag 
{\left(\rho^{(k)}_2\right)}^T$.} of the partially transposed density matrix 
${\left(\rho^{(k)}_2\right)}^T$.  It may be noted that negativity of partial transpose of any two-qubit system varies form  $0$ to  $0.5$. 
Here we adopt the convention in Ref.~\cite{neg} and redefine $N_{k_{2}}$ as   
\begin{equation} 
N_{k_{2}}=\vert\vert {(\rho^{(k)}_2)}^T \vert \vert-1 
\end{equation}
so that it takes values in the range $0$ to $1$  . 

We now proceed to evaluate the entanglement $N_{k_{1}}$ between any qubit of the state $\vert D_{N-k,\,k}\rangle$ with its remaining $N-1$-qubits, in terms of squared negativity of partial transpose. On recalling the result~\cite{neg} that negativity of partial transpose between a single qubit with the remaining qubits of any pure $N$-qubit state matches with the corresponding concurrence, we have  
\be
N_{k_{1}}=C_{k_{1}}=2\sqrt{\det\,\rho^{(k)}_1} 
\ee 
for the pure symmetric state $\vert D_{N-k,\,k}\rangle$.

As the pairwise entanglement  between all the $N-1$ pairs of qubits are equal, due to symmetry of the states belonging to Dicke-class, the monogamy inequality (See Eq. (\ref{moninqD})) turns out to be~\cite{neg},   
\be
\label{neg1}
N^2_{k_{1}}\geq  (N-1)N^2_{k_{2}} 
\ee
in terms of squared negativity of partial transpose.  
The bound on monogamy inequality or, equivalently the measure of residual entanglement of the states $\vert D_{N-k,\,k}\rangle$  is given by 
\be
\label{ntangle}
\xi^{(k)}_N=N^2_{k_{1}}-(N-1)N^2_{k_{2}}. 
\ee
where $\xi^{(k)}_N$ denotes the negativity tangle of the states $\vert D_{N-k,\,k}\rangle$ belonging to the Dicke-class. 
\begin{figure}[ht]
\includegraphics*[width=2.5in,keepaspectratio]{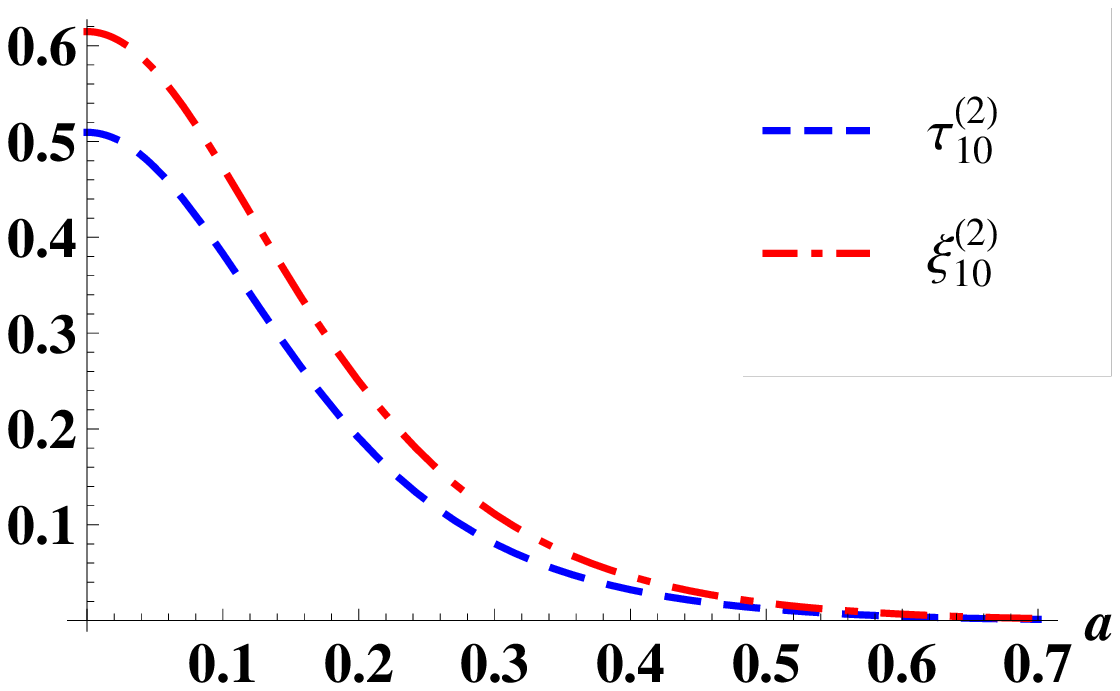}
\includegraphics*[width=2.5in,keepaspectratio]{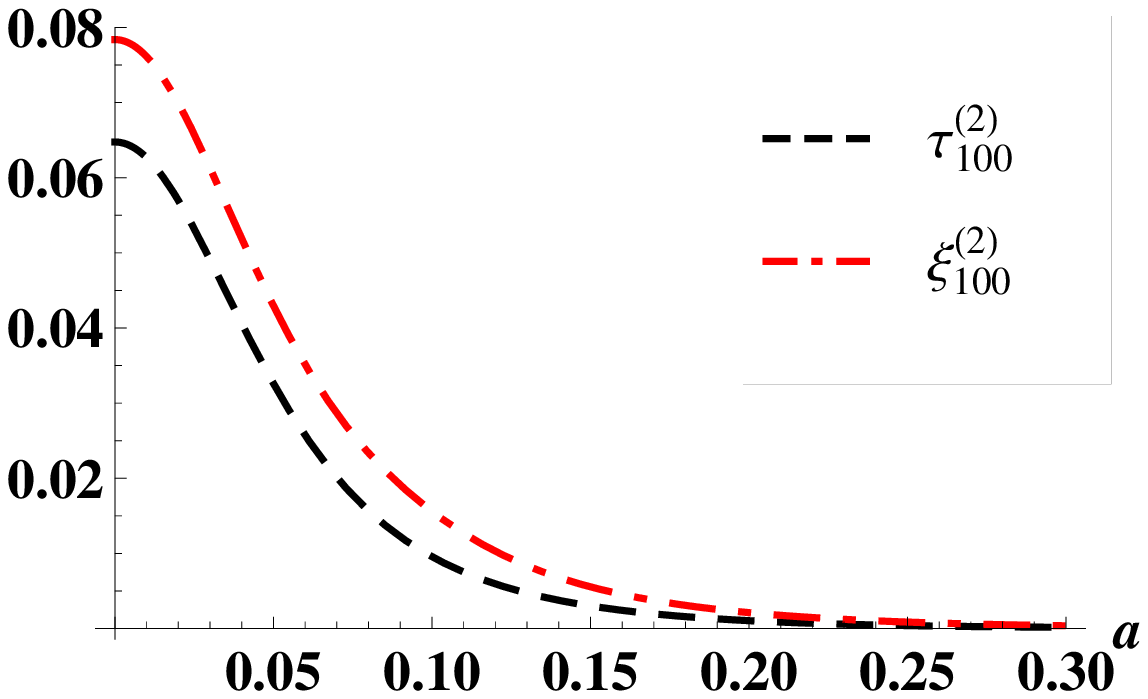}
 \caption{Comparison of concurrence tangle $\tau^{(2)}_N$ with negativity tangle $\xi^{(2)}_N$ for $N=10$ and $N=100$}  
\end{figure} 

\begin{figure}[ht]
\includegraphics* [width=2.5in,keepaspectratio]{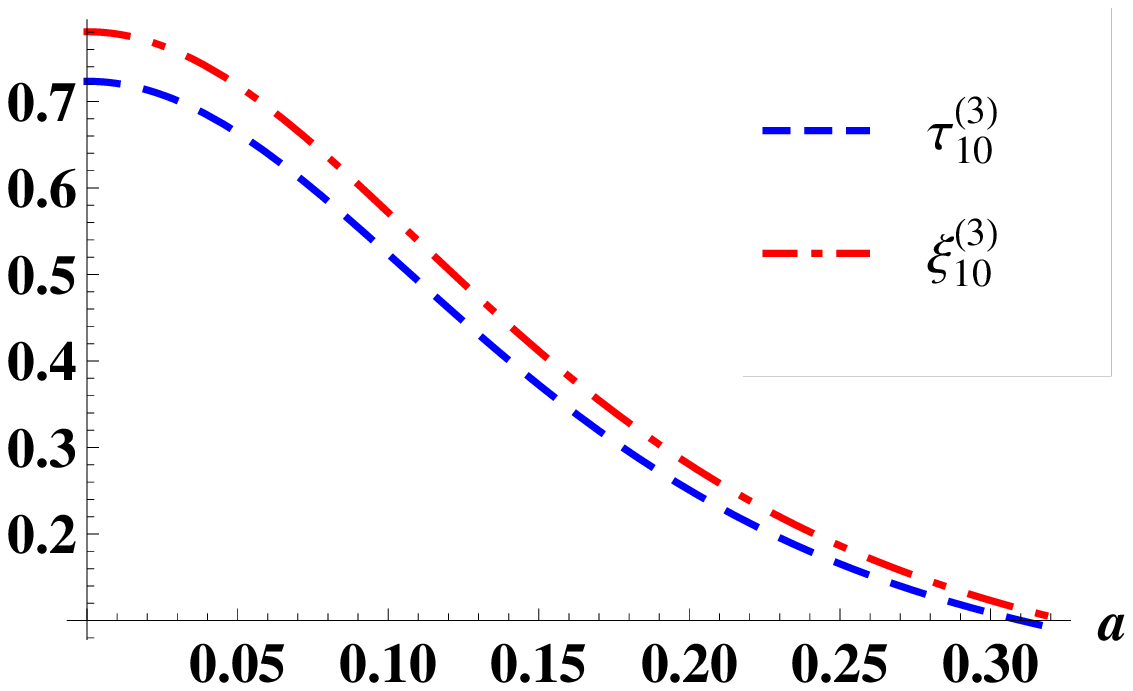}
\includegraphics* [width=2.5in,keepaspectratio]{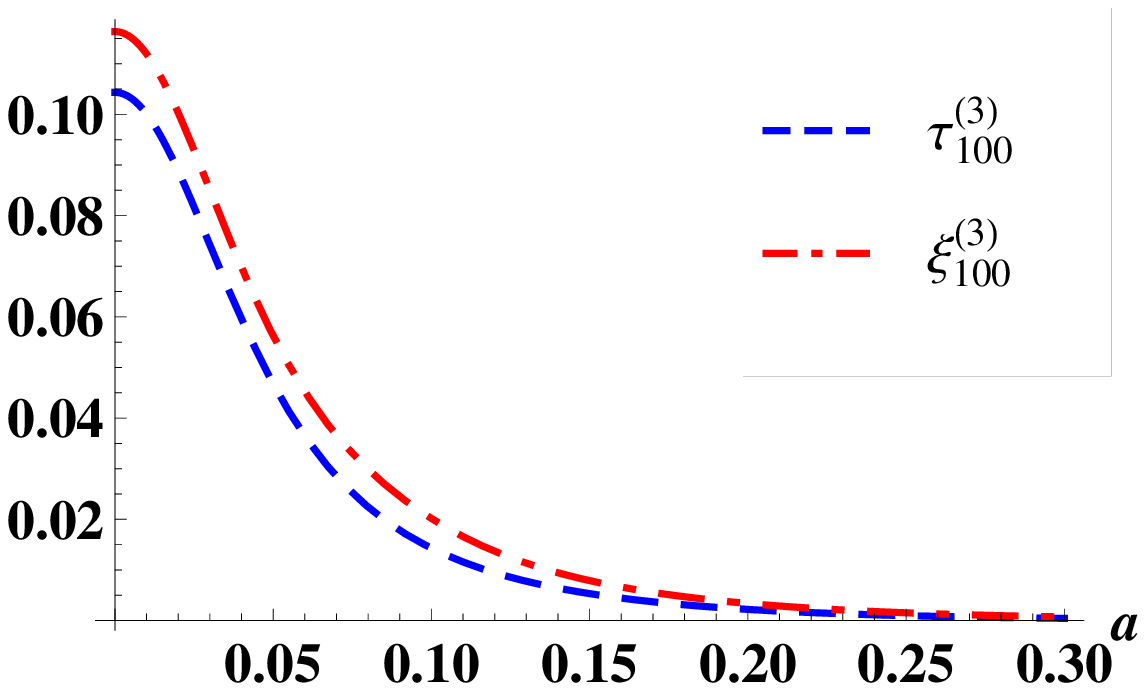}
 \caption{Comparison of concurrence tangle $\tau^{(3)}_N$ with negativity tangle $\xi^{(3)}_N$ for $N=10$ and $N=100$}  
\end{figure} 

\begin{figure}[ht]
\includegraphics* [width=4in,keepaspectratio]{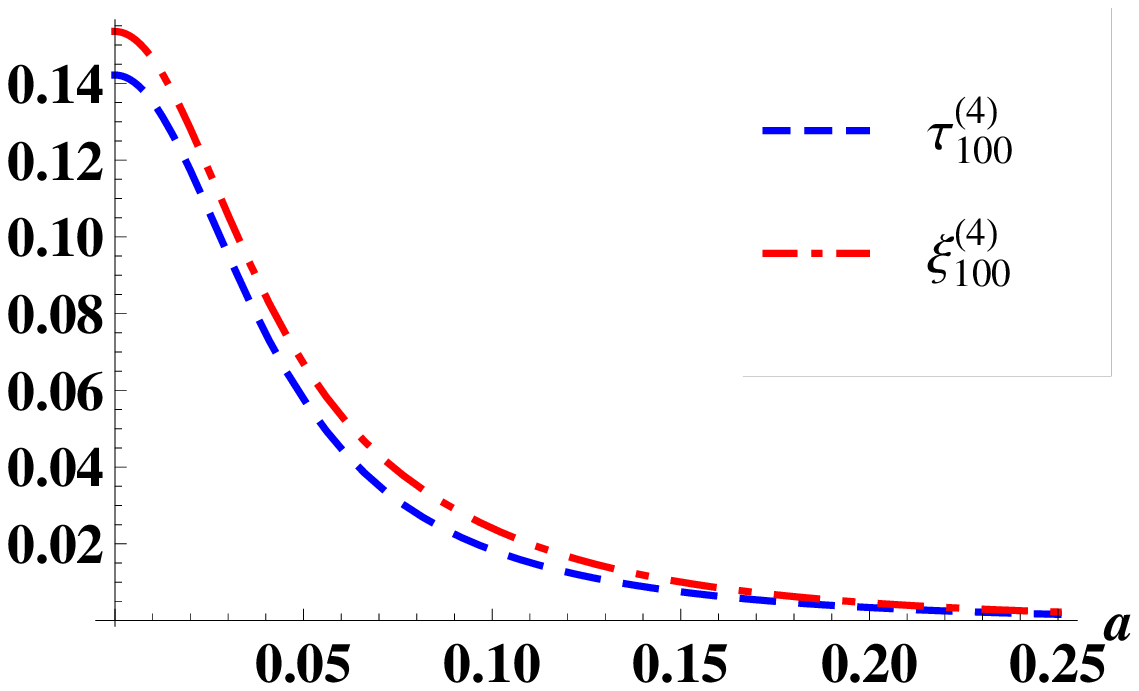}
 \caption{Comparison of concurrence tangle $\tau^{(4)}_N$ with negativity tangle $\xi^{(4)}_N$ for $N=100$}  
\end{figure} 

\begin{figure}[ht]
\includegraphics* [width=4in,keepaspectratio]{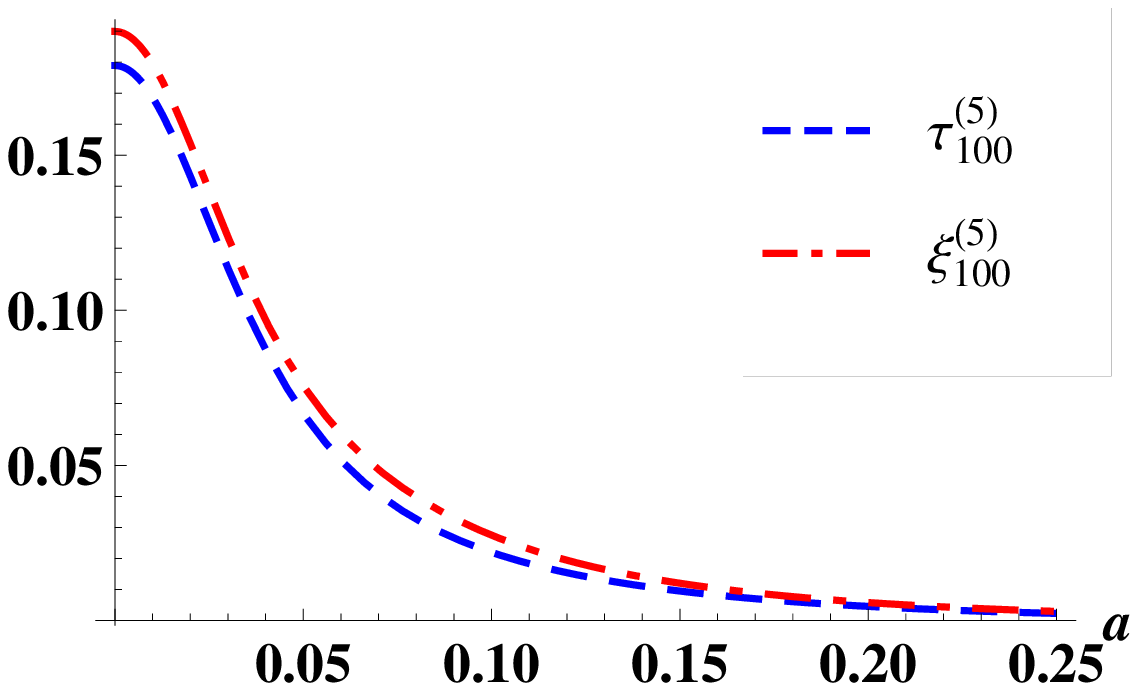}
\caption{Comparison of concurrence tangle $\tau^{(5)}_N$ with negativity tangle $\xi^{(5)}_N$ for $N=100$}  
\end{figure} 
 
In fact, for $k=1$, i.e., for  $\{{\cal D}_{N-1,1}\}$, the so-called W-class of states, concurrence tangle $\tau^{(1)}_N=0$ whereas negativity tangle $\xi^{(1)}_N$ is non-zero for all values of $N\geq 3$~\cite{pjg2}. This feature is seen in Fig. 9. 
  
\begin{figure}[ht]
\includegraphics* [width=4in,keepaspectratio]{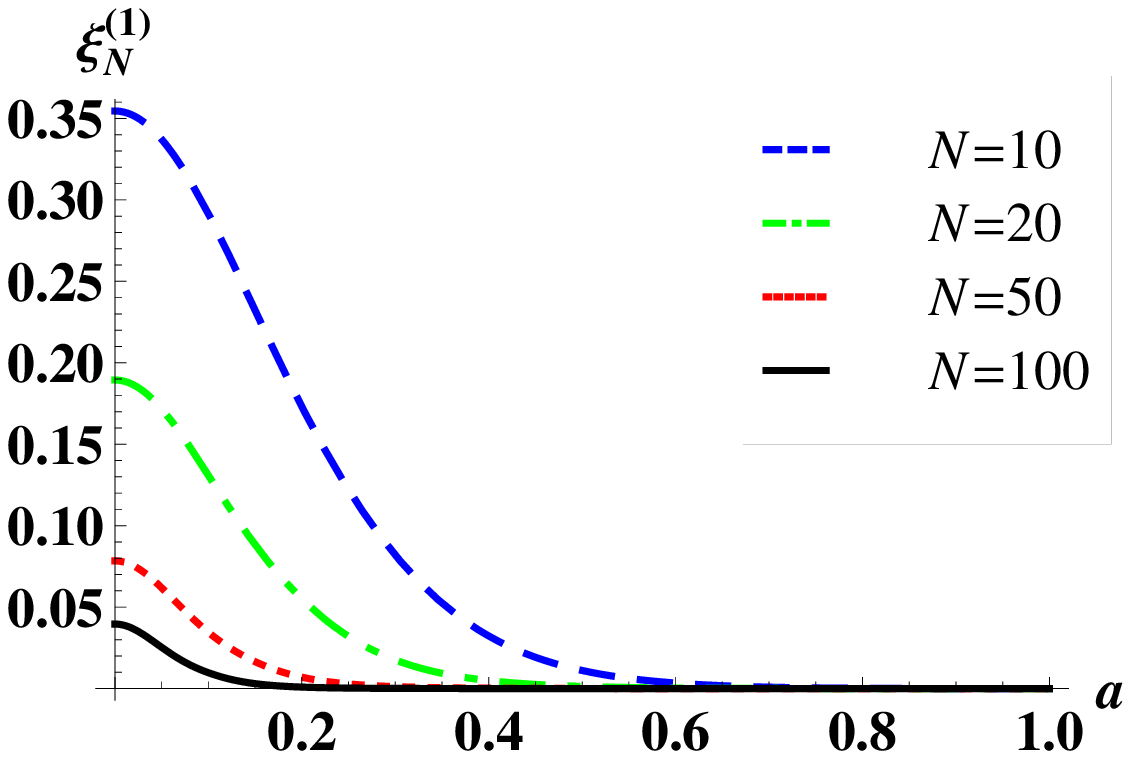}
 \caption{The variation of negativity tangle $\xi^{(1)}_N$ as a function of the parameter $a$ for different values of $N$.}  
\end{figure}

It may be seen, from Figs. 5 to 8, that the nature of variation of negativity tangle $\xi^{(k)}_N$ is quite similar to that of concurrence tangle $\tau^{(k)}_N$ (See Figs. 1 to 4) for each value of $k$ ($k=2,\,3,\,4,\,5$) and $N$. We also notice that
$\xi^{(k)}_N\geq \tau^{(k)}_N$ (See Figs. 5 to 9). Thus, the bound on monogamy inequality in terms of squared negativity of partial transpose exceeds the bound in terms of squared concurrence. In other words, negativity tangle $\xi^{(k)}_N$ is greater than or equal to concurrence tangle $\tau^{(k)}_N$, quite in accordance with the result obtained in Ref.\cite{neg} for $N$ qubit states of the W-class. We therefore conclude that for the entire Dicke-class $\{{\cal D}_{N-1,1}\}$,  containing all possible SLOCC inequivalent classes of states 
$\vert D_{N-k,\,k}\rangle$, the nature of variation of the bound on monogamy inequality, expressed in squared concurrence, squared negativity of partial transpose is in agreement with that observed for W-class of states~\cite{pjg2}. 

\section{Conclusion} 
In this article, we have analyzed the monogamous nature of $N$-qubit pure symmetric states with two distinct Majorana spinors--the so-called 
Dicke-class of states, using squared concurrence and squared negativity of partial transpose as measures of two-qubit entanglement.  
Towards this end, we have made use of Majorana geometric representation to obtain a simplified, one parameter structure of Dicke-class of states. The familiar angular momentum algebra relating to addition of angular momentum of two spin systems, 
enables us to partition this simplified form of the generalized Dicke states into two of its subsystems.
Using this partitioning we obtain the general form of two-qubit and single-qubit density matrices of Dicke-class of states. The reduced density matrices so obtained allow us to determine the pairwise entanglement and residual entanglement (which goes beyond the pairwise engtanglement) in the system. 

The bound on monogamy relation, using squared concurrence as measure of entanglement, is analyzed for all SLOCC inequivalent families of states belonging to Dicke-class. Moreover our results reveal that, among the several inequivalent families, corresponding to different degeneracy configuration of the two spinors, the states belonging to the family having equal distribution of its two spinors are found to have less shareability amongst its qubits(more monogamous), thereby possessing larger residual entanglement. 

The monogamy inequality in terms of squared negativity of partial transpose is also analyzed and it is seen that for the entire Dicke-class of states, the residual entanglement quantified through negativity tangle exceeds the one quantified through concurrence tangle. Our results confirm that both the measures of entanglement are analogous in establishing monogamous nature in Dicke-class of states and either of the two measures can be chosen for the evaluation of their residual entanglement.  

Summarizing, our work accomplishes the task of analyzing the monogamous nature of the entire one parameter family of pure symmetric multiqubit states characterized by two-distinct spinors, for the first time. The results of this work opens up avenues for potential applications of Dicke-class of states in the field of quantum information processing, as limited shareability of entanglement among subsystems/monogamous nature happens to be one of the highlighted {\emph{non-classical}} features of multiparty quantum systems.

\section*{Acknowledgements}
KSA would like to thank the University Grants Commission for providing a BSR-RFSMS fellowship during the present work. ARU acknowledges the support of UGC MRP (Ref. MRP-MAJOR-PHYS-2013-29318).

\end{document}